\documentclass{article}

\usepackage{soul}
\usepackage{ccaption}
\usepackage{color}
\usepackage{pifont}
\usepackage{mathptmx}
\usepackage{amsmath, amsfonts, amssymb, mathrsfs}
\usepackage{graphicx}

\DeclareMathOperator*{\argmax}{arg\,max}
\DeclareMathOperator*{\argmin}{arg\,min}
\usepackage{multirow}
\usepackage{multicol}
\usepackage[figuresright]{rotating}
\usepackage[round]{natbib}

\title{Estimation and Evaluation of the Resource-Constrained Optimal Dynamic Treatment Rule: An Application to HIV Care Retention}
\author{Lina M. Montoya \thanks{School of Data Science and Society and Department of Biostatistics, University of North Carolina at Chapel Hill} 
\and Elvin Geng \thanks{Division of Infectious Diseases, Washington University in St. Louis School of Medicine} 
\and Harriet F. Adhiambo \thanks{Department of Child, Family, and Population Health Nursing, University of Washington} 
\and Maya Petersen \thanks{University of California, Berkeley}}
\date{}

\begin{document}
\maketitle

\begin{abstract}
The optimal strategy for deploying a treatment in a population may recommend giving all in the population that treatment. Such a strategy may not be feasible, especially in resource-limited settings. One approach for determining how to allocate a treatment in such settings is the resource-constrained optimal dynamic treatment rule (RC ODTR) SuperLearner algorithm, developed by Luedtke and van der Laan. In this paper, we describe this algorithm, offer various novel approaches for presenting the RC ODTR and its value in terms of benefit and cost, and provide practical guidance on implementing the algorithm (including software). In particular, we apply this method to the Adaptive Strategies for Preventing and Treating Lapses of Retention in HIV care (NCT02338739) trial to determine how to best allocate conditional cash transfers (CCTs) for increasing HIV care adherence given varying constraints on the proportion of people who can receive CCTs in the population, providing one of the few applied illustrations of this method and novel substantive findings. We find that there are clinical and monetary benefits to deploying CCTs to a small percent (e.g., 10\%) of the population compared to administering the care standard for all; however, results suggest that these incremental benefits are only due to the loosening of constraints, rather than a presence of treatment effect heterogeneity strong enough to drive a more efficient and effective constrained allocation approach.
\end{abstract}

\section{Introduction}
\label{s:intro}

The goal of precision health is to shift scientific questions from ``does a treatment work, on average?" to ``for whom does a treatment work?", to ultimately learn the most efficient and effective ways of administering treatments. Approaches for answering the latter question aim to identify differential treatment responses, and thus learn how to administer treatment  only to those persons who are likely to benefit from it. Traditional empirical methods for identifying such persons include, for example, subgroup analyses for understanding treatment effect heterogeneity within (ideally) \textit{a priori} specified subgroups. More recently, approaches, including those leveraging machine learning, have been proposed to estimate optimal dynamic treatment rules \citep{murphy2003} (i.e. policies) – algorithms that input an individual's covariate information and output the treatment decision for that individual that has the best expected outcome (see, e.g., \cite{kosorok2019}, for a review of estimation approaches for these algorithms). Such approaches are particularly useful in settings in which a treatment that is beneficial on average may nonetheless be harmful to some subset of the population, or when multiple treatment options are available and persons differ in the option to which they are most likely to respond.

It may be the case, however, that resources are limited such that a treatment cannot be given to more than a certain proportion of the population. In such a scenario, improved understanding of treatment effect heterogeneity is crucial for effective resource allocation, even in the simple scenario of a single candidate treatment that is beneficial on average overall and conditional on all measured covariates. Consider the following real-world example: among adults living with HIV in sub-Saharan Africa, it was found that, on average, conditional cash transfers (CCTs) helped persons living with HIV (PWH) stay engaged in HIV care, relative to the standard-of-care \citep{geng2023adaptive}. Here, it may be that CCTs do not harm anyone; everyone has an expected positive effect of CCTs, conditional on measured covariates. In this scenario, the optimal dynamic rule would say: ``give CCT treatment to everyone," regardless of  covariate history. However, it may not be financially feasible to give CCTs to all persons in this population to maintain their HIV care. In other words, in this situation, resources constrain the proportion of people who can receive CCTs over a period of time. Under this constraint, a treatment will have the greatest benefit if it is allocated to those persons most likely to benefit. Informally, our precision health question shifts further, from ``for whom does a treatment work?" to ``for whom does a treatment work \emph{the most}?"

Several methodological approaches exist for selectively administering treatment, subject to resource constraints (e.g., \cite{caniglia2021estimating, sarvet2024optimal, xu2024optimal}). One approach aims to estimate the so-called resource-constrained optimal dynamic treatment rule (RC ODTR), proposed by \cite{luedtke2016statistical}. The RC ODTR is the optimal dynamic treatment rule that respects the constraint that only a proportion of the population is able to receive the treatment. An unconstrained optimal rule could be defined as a function of the conditional average treatment effect (CATE): if a person's CATE is greater than 0 (i.e., positive treatment effect), treat that person; else, do not treat that person. The RC ODTR changes the original optimal rule's threshold to one that is greater than 0, thereby treating only those who have a treatment effect higher than some cutoff determined by the constraint on the proportion of people who can receive treatment. Further, under this algorithm, if two or more people are ``tied" for resources, the rule becomes stochastic (i.e., those people are given a probability of treatment, as opposed to a deterministic, binary treatment decision). This can occur if, for example, the optimal dynamic treatment rule is a function of only discrete covariates.

Although the RC ODTR algorithm has been described in theoretical detail (e.g., \cite{luedtke2016statistical}) and presented as part of a larger clinical trial analysis (e.g., \cite{kessler2022individualized}), to our knowledge, there are no articles in the literature describing and illustrating this method in the context of an applied example. Filling this gap may be useful for analysts who would like to apply the RC ODTR algorithm to their data. Thus, in this article, we present an applied example of the RC ODTR on data generated from the Adaptive Strategies for Preventing and Treating Lapses of Retention in HIV Care (ADAPT-R; NCT02338739) trial. Specifically, we apply the RC ODTR on ADAPT-R data to understand which individuals are most likely to benefit from CCTs under a constraint on the proportion of people who can receive CCTs in the population, and quantify the clinical loss incurred by use of a constrained vs. unconstrained rule. We also build on prior work by illustrating how the impact of a varying resource constraint level can be summarized using a causal estimand defined with a marginal structural model (MSM), and use results from this to determine the utility of using an RC ODTR to determine, among a population of persons all expected to benefit from a treatment, who should receive a resource-constrained intervention, compared to implementing the resource constraint at random. Finally, we show how to construct an incremental cost-effectiveness ratio (ICER) illustrating the cost effectiveness of implementing such rules, as a function of different resource constraint levels. In presenting these results, we hope to show practical ways of using and evaluating the RC ODTR to understand how to selectively administer treatments to only those persons most likely to benefit from them, and any costs to doing so. These analyses are additionally accompanied by an R package (https://github.com/lmmontoya/SL.ODTR).

The article is organized as follows: in Section 2, we describe the data and specify a causal model on the data generating process that gave rise to our data, including a description of the ADAPT-R example. In Section 3, we present preliminary evidence of treatment effect heterogeneity. In Section 4, we give an overview of the RC ODTR algorithm. In Section 5, we discuss identification, estimation, and inference of RC ODTR-based parameters (i.e., causal estimands). In Section 6, we apply these methods to the ADAPT-R study. We close with a discussion.

\section{Data and Causal Model}

We consider point-treatment data in which patients have baseline covariates $W \in \mathcal{W}$, a binary treatment variable $A \in \{0,1\}$, and an outcome variable $Y \in \mathbb{R}$ measured at the end of the study. We assume these data are generated by a process compatible with the following structural causal model (SCM) $\mathcal{M}^F$ \citep{pearl2000}:
\begin{align*}
    W& =f_W (U_W )\\
    A&=f_A (W,U_A)\\
    Y&=f_Y (W,A,U_Y). 
\end{align*}
Here, $U = (U_W, U_A, U_Y) \sim P_U$ are unmeasured exogenous variables, $f = (f_W, f_A, f_Y)$ are structural equations, and the random variables in this SCM follow the joint distribution $P_{U,X}$.  Additionally, $f_A (W,U_A) = \mathbb{I}[U_A < g(1|W)]$, where $U_A \sim Uniform(0,1)$ and $g(A|W) = Pr_{P_{U,X}}(A|W)$; in other words, $A \sim Bernoulli(p = g(1|W))$. As in the ADAPT-R trial (described in the following section), it may be known that the treatment was was randomized with equal probability to each arm; in that case, it is known that $g(1|W) = 0.5$, $U_A \perp U_W$ and $U_A \perp U_Y$, and the above structural causal model would state that $Y$ may be affected by both $W$ and $A$, but that $W$ does not affect $A$. The results presented here apply, however, to both randomized trial and observational data settings.

\subsection{The ADAPT-R Trial}

Many people living with HIV do not stay in care over time, and missed clinic visits can compromise HIV care success \citep{gengestimation}. The ADAPT-R trial was carried out among people living with HIV and initiating antiretroviral treatment (ART) in the Nyanza region of Kenya to evaluate behavioral interventions to optimize successful HIV care outcomes \citep{geng2023adaptive}. Nested within ADAPT-R were several randomized controlled trials (RCTs). In one, 1,189 persons were randomized to one of two initial interventions to prevent a lapse in care: either conditional cash transfers (CCTs) in the amount of 400 Kenyan shillings for on-time clinic visits or standard of care (SOC) education and counseling. The primary outcome was treatment success, defined as having HIV viral suppression at end of year 1 \textit{and} remaining in care throughout year 1 after baseline randomization. The following baseline covariate information was also collected on each participant: sex, age, CD4+ T cell count, depression, social support, food insecurity, presence of treatment support, distance participant walks to clinic, wealth index, total hours spent on any work, total income from any source in the last month, total discretionary spending in the last month, total fixed spending in the last month, work for wages (casual work), farming in household, self-employed work, BMI, pregnancy, education level, occupation, marital status, and WHO HIV disease stage. Primary trial analyses showed that CCTs were beneficial, on average -- there was a 9.9\% increase (95\% CI: 4.6\%, 15.2\%) in the probability of treatment success under a CCT intervention versus SOC.

\section{Preliminary Evidence of Treatment Effect Heterogeneity}

Patients who experience a care lapse have a diversity of barriers to retention \citep{gengestimation}, demonstrating the need for personalized strategies to increase successful HIV care outcomes. For example, it may be the case that CCTs provide an effective incentive to stay in care for some persons, but have no effect \citep{giordano2024experiences, mccoy2017cash, el2017financial, el2019brief}, or are detrimental to care engagement, among others. Thus, we estimated the optimal (unconstrained) dynamic treatment rule on the ADAPT-R data to determine which approach to allocating CCTs versus SOC adults living with HIV would result in the highest overall treatment success. 

\subsection{The Unconstrained Optimal Dynamic Treatment Rule}\label{odtr_noRC}

The unconstrained optimal dynamic treatment rule has been described in detail elsewhere \citep{murphy2003, kosorok2019, robins2004, montoya2022optimal}. Briefly, a dynamic treatment rule $d$ is defined as a function that takes in measured covariates (or a decision function of those measured covariates, $V$, although here we will only consider cases where $V = W$ and thus use $W$ as input) and outputs a treatment decision $d:\mathcal{W} \rightarrow \{0,1\}$. The value of a rule $d$ is $\mathbb{E}_{P_{U,X} } [Y_d]$. The optimal \textit{dynamic} treatment rule $d^*$ is the rule, among the set of all dynamic treatment rules $\mathcal{D}$, that maximizes expected counterfactual outcomes (assuming larger outcomes are better). That is, $d^*(W) \in \argmax_{d\in \mathcal{D}} \mathbb{E}_{P_{U,X} } [Y_d]$. Assuming that assigning no treatment is preferable in the absence of a treatment effect, one can equivalently define the optimal dynamic treatment rule as a function of the conditional average treatment effect (CATE) $\mathbb{E}_{P_{U,X}}[Y_1-Y_0|W]$; that is, $d^*(W)= \mathbb{I}\Big{[}\mathbb{E}_{P_{U,X}}[Y_1-Y_0|W] > 0\Big{]}$. Under the point-treatment randomization and positivity (e.g., \cite{petersen2012diagnosing}) assumptions, these causal parameters are identified as statistical parameters, or functions of the observed data distribution $P_0$ (an element of the statistical model $\mathcal{M}$). For example, the CATE is identified as the so-called ``blip function" $B_0(W) = \mathbb{E}_0[Y|A=1,W] - \mathbb{E}_0[Y|A=0,W]$, which implies the identified optimal dynamic treatment rule: $d_0^*(W) = \mathbb{I}[B_0(W)>0]$. 

\begin{figure}[h]
    \centering
    \includegraphics[scale = .5]{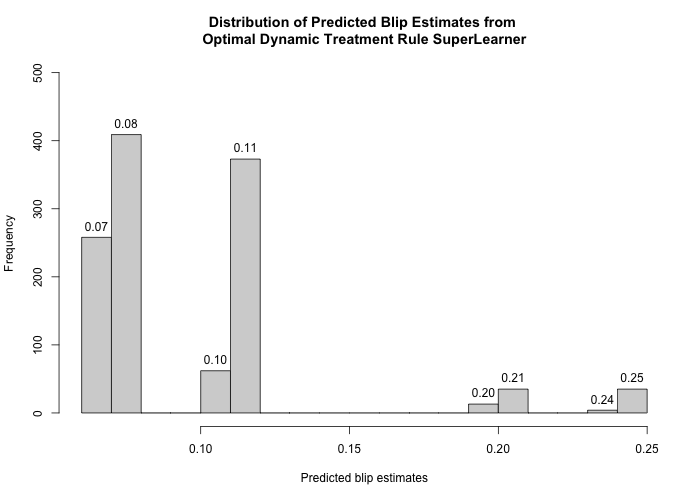}
    \caption{Distribution of predicted conditional average treatment effect/blip estimates from the optimal dynamic treatment rule SuperLearner. The frequencies are divided into eight discrete categories (bars) because the algorithm allocated all coefficient weights to three generalized linear models, each with a binary predictor variable. The numbers on top of each bar are the exact blip estimates corresponding to that covariate category. All blip values are positive, implying that the estimated optimal dynamic treatment rule recommended treatment (conditional cash transfers) to all.}
    \label{blipfig}
\end{figure}

We can estimate the blip function and thus optimal dynamic treatment rule using the Optimal Dynamic Treatment Rule SuperLearner \citep{montoya2022optimal, luedtkeSLODTR}. Using a diverse library of candidate algorithms consisting of simple, parametric models (univariate generalized linear models [GLMs] with each covariate) and other/more flexible machine learning algorithms (\texttt{SL.mean},  \texttt{SL.glm}, \texttt{SL.bayesglm}, \texttt{SL.stepAIC}); a blip-based metalearner; mean squared error (MSE) as a risk function; and 10-fold cross validation, the estimated optimal rule for administering CCTs for the ADAPT-R population recommended giving CCTs to all. The estimated weights of this SuperLearner can be found in Appendix A. The distribution of the estimated CATE (i.e., the estimated identified blip function) is illustrated in Figure \ref{blipfig} and Table \ref{table1}, in which all estimated values are positive. We note that because the unconstrained optimal dynamic treatment rule SuperLearner placed all weight on three simple GLMs that were each a function of one binary variable (whether a person walked more than 5 kilometers to clinic, was self-employed, and worked for wages), the distribution of the estimated blip is discretely divided into eight possible categories.

\begin{table}[ht]
\centering
\begin{tabular}{lll}
  \hline
Blip estimate & N (\%) & Covariate category \\ 
  \hline
0.07 & 258 (21.70\%) & Yes wage work, No self-employed, No walk $>$ 5 km \\ 
  0.08 & 409 (34.40\%) & No wage work, No self-employed, No walk $>$ 5 km \\ 
  0.10 & 62 (5.21\%) & Yes wage work, Yes self-employed, No walk $>$ 5 km \\ 
  0.11 & 373 (31.37\%) & No wage work, Yes self-employed, No walk $>$ 5 km \\ 
  0.20 & 13 (1.09\%) & Yes wage work, No self-employed, Yes walk $>$ 5 km \\ 
  0.21 & 35 (2.94\%) & No wage work, No self-employed, Yes walk $>$ 5 km \\ 
  0.24 & 4 (0.34\%) & Yes wage work, Yes self-employed, Yes walk $>$ 5 km \\ 
  0.25 & 35 (2.94\%) & No wage work, Yes self-employed, Yes walk $>$ 5 km \\ 
   \hline \\ \\
\end{tabular}
\caption{Distribution of the blip/CATE estimated by SuperLearner. Because the algorithm only returned non-zero coefficients on three binary variables (wage work, self-employment, walk $>$ 5 km), the support of the estimated blip distribution only takes on eight values, listed in the ``Covariate category" column.} 
\label{table1}
\end{table}

\subsection{Subgroup Analysis}

Based on results from the optimal dynamic treatment rule SuperLearner, the most effective rule for maximizing treatment success would recommend: ``give CCTs to all persons." In practice, however, giving CCTs to all may not be feasible, and constraints on CCT administration can be even more pronounced in resource-limited settings. Indeed, in ADAPT-R, despite CCTs being most effective, CCT-based interventions were also shown to be the most costly \citep{montoya2025cost}. One way to optimize the impact of deploying an effective, but costly, intervention under resource constraints is to selectively administer the intervention to patient subgroups who are \textit{most likely} to benefit.

\begin{figure}[h]
    \centering
    \includegraphics[scale = .6]{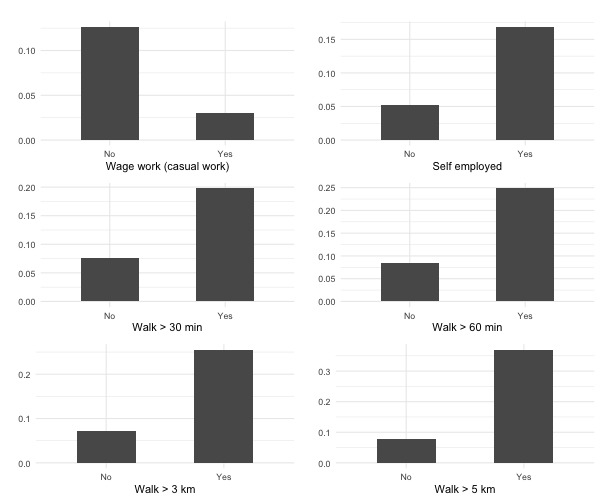}
    \caption{Subgroup plots for each of the significant effect modifiers in the ADAPT-R data. The x-axis for each of the plots is the different levels of the covariates; the y-axis is the estimated effect of CCTs versus SOC, in that covariate subgroup.}
    \label{subgroupfig}
\end{figure}

Subgroup analyses -- that is, estimating the unadjusted difference in probability of treatment success between those who received CCT versus SOC, within each covariate group level -- we move closer to understanding which persons are most likely to benefit. In Figure \ref{subgroupfig}, we show subgroup plots for each covariate in which a likelihood ratio test showed a significant ($p < 0.1$) interaction on the linear regression of CCT treatment, covariate, and treatment by covariate interaction. These exploratory analyses provide insights into the subgroups most likely to benefit from CCTs -- for example, those who work for wages (i.e., partake in casual work), are self-employed, walk $>$ 30 minutes to clinic, or walk $>$ 3 kilometers to clinic. However, these subgroups were not pre-specified \textit{a priori}; appropriate interpretation of these analyses requires correction for multiple testing. Further, these analyses consider only subgroups defined by single covariates, without considering those defined by covariate interactions. The goal is then to take a wide range of participant characteristics and flexibly combine them to identify those who most benefit from CCTs to preferentially allocate CCTs to those participants.

\section{The Resource-Constrained Optimal Dynamic Treatment Rule}

With the RC ODTR, one can estimate the optimal strategy for allocating treatment, subject to constraints on the proportion of people who can receive treatment in the population \citep{luedtke2016statistical}. The RC ODTR does this by identifying subgroups defined by $W$ who are most likely to benefit from treatment. Following, we provide an overview of the RC ODTR, including an illustration of how stochastic rules are used for allocation in the case of ``ties" (i.e. persons with equivalent CATE estimates), and how to evaluate a given RC ODTR.

\subsection{The Resource-Constrained Optimal Deterministic Rule}

The deterministic RC ODTR, which we denote as $d^*_{\kappa}(W)$, is defined as the deterministic rule $d$ that solves the following constrained optimization problem:
\begin{align}
\text{Maximize } \mathbb{E}_{P_{U,X} } [Y_d] \text{ subject to } \mathbb{E}_{P_{U,X}}[d(W)] \leq \kappa,
\label{definition}
\end{align}
where $\kappa$ is the investigator-specified maximum proportion of the population who can receive treatment $A=1$. In the ADAPT-R case, $\kappa$ is the maximum proportion of people we, as investigators, allow to receive CCTs in the population. Thus, the deterministic RC ODTR is the deterministic rule that yields the highest value, under the constraint that the proportion of people who are treated in the population $\mathbb{E}_{P_{U,X}}[d(W)]$ is less than or equal to $\kappa$. The deterministic RC ODTR can be equivalently be defined as a function of the CATE, i.e., 
\begin{align}
    d^*_{\kappa}(W) &= \mathbb{I}\Big{[}\mathbb{E}_{P_{U,X}}[Y_1-Y_0|W] > \tau_{P_{U,X}}\Big{]},
    \label{detrule_causal}
\end{align}
where $\tau_{P_{U,X}} = \max\{\eta_{P_{U,X}},0\}$, $\eta_{P_{U,X}} = \inf\{\tau : S_{P_{U,X}}(\tau) \leq \kappa\}$, and $S_{P_{U,X}}(\tau) = Pr(\mathbb{E}_{P_{U,X}}[Y_1-Y_0|W] > \tau)$. Intuitively, the optimal dynamic treatment rule becomes resource-constrained by increasing the original threshold of 0 on the unconstrained rule to $\tau_{P_{U,X}}$ (i.e., instead of treating those for whom CATE is larger than 0 as in the unconstrained rule, treat those for whom the CATE is larger than $\tau_{P_{U,X}}$), such that the proportion of people treated under that new threshold is less than or equal to the investigator-imposed constraint $\kappa$. In this way, if the proportion of people who have a treatment effect greater than 0 is less than $\kappa$, then the threshold will remain at $\tau_{P_{U,X}} = 0$, meaning that then the RC ODTR $d^*_{\kappa}(W)$ is equal to the unconstrained optimal rule $d^*(W)$.

Importantly, we note that (\ref{detrule_causal}) only holds as the maximizer for (\ref{definition}) if $Pr(\mathbb{E}_{P_{U,X}}[Y_1-Y_0|W] = \tau_{P_{U,X}}) = 0$; that is, that no covariate profile $W=w$ has a CATE at exactly the new threshold, $\tau_{P_{U,X}}$. This can occur when the optimal rule is a function of discrete covariates, and therefore the distribution of the CATE that defines that optimal rule is not continuous. For example, consider the estimated optimal dynamic treatment rule in ADAPT-R, which is based on three binary variables. As shown in Figure \ref{blipfig}, the estimated CATE distribution is split into $2^3 = 8$ categories (where 3 is the number of variables in the CATE and 2 is the support for each of those variables). Those in the lowest of the eight CATE categories (corresponding to an estimated CATE of 0.07) -- those who work for wages, but do not walk more than 5 kilometers or are self-employed -- comprise the first 21.70\% of the CATE distribution. If the constraint $\kappa$ allows CCT deployment to at most between 100\% - 21.70\% = 78.30\% and 100\% of the population, it is unclear how to allocate CCTs among those with the lowest CATE values. Said another way, if, for example, $\kappa = 0.9$, then the segment of the population who work for wages, do not walk more than 5 kilometers, and are not self-employed could not all receive CCTs under such a constraint, and thus would be tied for any CCTs that remain after preferentially allocating them to the other seven subgroups with a higher CATE.






\subsection{The Resource-Constrained Optimal Stochastic Rule}

The resource-constrained optimal stochastic rule is one approach for dealing with the scenario in which $Pr(\mathbb{E}_{P_{U,X}}[Y_1-Y_0|W] = \tau_{P_{U,X}}) = 0$, where there may be potential ``ties" for determining how to allocate treatment among those with the same CATE. Consider a stochastic rule $\tilde{g}(W)$, which we define as a function that takes in measured covariates $W$ (or a decision function of those measured covariates, as above) and outputs a probability of receiving treatment, i.e., $\tilde{g} : \mathcal{W} \rightarrow [0,1]$. 
 
We can intervene on the above SCM to derive counterfactual outcomes $Y_{\tilde{g}}$ under a given stochastic intervention by setting $A$ equal to $a$ with probability $\tilde{g}(a|W)$, which we denote $A^*$, so that $A^* \sim Bernoulli(p = \tilde{g}(1|W)$. Then,  $Y_{\tilde{g}}$ can be interpreted as the counterfactual outcome for a patient if their treatment $A$ were assigned using the stochastic treatment rule $\tilde{g}$. The value of a stochastic rule $\tilde{g}$ is the expected outcome under $\tilde{g}$, or $\mathbb{E}_{P_{U,X}} [Y_{\tilde{g}}]$. We note that the unconstrained optimal stochastic rule $\tilde{g}^*$ is equivalent to the unconstrained optimal deterministic rule $d^*$ \citep{montoya2020estimation}.

The stochastic RC ODTR, which we denote as $\tilde{g}^*_{\kappa}(W)$, is defined as the stochastic rule $\tilde{g}$ that solves the following constrained optimization problem:
\begin{align*}
\text{Maximize } \mathbb{E}_{P_{U,X} } [Y_{\tilde{g}}] \text{ subject to } \mathbb{E}_{P_{U,X}}[\tilde{g}(W)] \leq \kappa.
\end{align*}
Similar to the deterministic RC ODTR, the stochastic RC ODTR can equivalently be defined as a function of the CATE, i.e., 
\begin{align*}
 \tilde{g}^*_{\kappa}(1|W) &=
  \begin{cases}
   \frac{\kappa - S_{P_{U,X}}(\tau_{P_{U,X}})}{Pr(\mathbb{E}_{P_{U,X}}[Y_1-Y_0|W] = \tau_{P_{U,X}})},        & \text{if } \mathbb{E}_{P_{U,X}}[Y_1-Y_0|W] = \tau_{P_{U,X}} \text{ and } \tau_{P_{U,X}} > 0 \\
   \mathbb{I}[\mathbb{E}_{P_{U,X}}[Y_1-Y_0|W] > \tau_{P_{U,X}}],        & \text{otherwise.}
  \end{cases}
\end{align*}
In other words, treatment is assigned according to the deterministic RC ODTR as long as this results in unabiguous assignment.  For the remaining $W=w$ for whom there are ``ties" for determining how to allocate treatment among those with the same CATE, a probability of treatment is given such that the constraints are respected (i.e., any treatments that remain available for allocation are randomly assigned among the persons with the same CATE).

\subsection{Evaluating the Resource-Constrained Optimal Dynamic Treatment Rule}

It may not only be of interest to obtain the RC ODTR, but also to evaluate it. For an investigator-specified constraint $\kappa$, a counterfactual outcome under the $\kappa$-specific RC ODTR is denoted $Y_{\tilde{g}^*_{\kappa}}$, which is generated by setting $A$ equal to $a$ in the structural equations with probability $\tilde{g}^*_{\kappa}(a|W)$, which we denote $\tilde{A}^*$. Then, the value under a $\kappa$-specific RC ODTR is $\Psi^F_{\kappa}(P_{U,X}) \equiv \mathbb{E}[Y_{\tilde{g}^*_{\kappa}}]$, where $\Psi^F:\mathcal{M}^F \rightarrow \mathbb{R}$. In ADAPT-R, for example, this would translate to the probability of treatment success under the RC ODTR in which at most $\kappa \times 100 \%$ of the population can receive CCTs. This causal quantity can be compared to the value of any other rule $d$ (for example, the value under the optimal dynamic treatment rule, $\mathbb{E}_{P_{U,X}}[Y_{d^*}]$, to understand any dropoff in the probability of treatment success by administering CCTs under a constrained rule versus the optimal rule with no constraints). We note that $\Psi^F_{\kappa=0}(P_{U,X}) = \mathbb{E}_{P_{U,X}}[Y_0]$, and in the case where treatment is likely to benefit all, as in ADAPT-R, $\Psi^F_{\kappa=1}(P_{U,X}) = \mathbb{E}_{P_{U,X}}[Y_1]$.

It may also be of interest to summarize how the expected counterfactual outcome (e.g., probability of treatment success) varies as a function of incremental constraints. Such a summary can be expressed via a causal ``constraint-response" curve $\left\{\mathbb{E}_{P_{U,X}}\left[\Psi^F_{\kappa}(P_{U,X})\right] : \kappa \in \bar{\kappa}\right\}$ for a subset of constraints $\bar{\kappa} \subseteq [0,1]$. The location of the constraint-response curve relative to the straight line between $\mathbb{E}_{P_{U,X}}[Y_0]$ and $\mathbb{E}_{P_{U,X}}[Y_1]$ may be of particular interest, because the latter line represents the expected outcome if a given constraint were implemented randomly without any covariate information. In cases where treatment is beneficial for all possible covariate values  (as in ADAPT-R), the constraint-response curve will be monotonically increasing and lie either on or mostly above the straight line connecting $\mathbb{E}_{P_{U,X}}[Y_0]$ and $\mathbb{E}_{P_{U,X}}[Y_1]$. If all points of the constraint-response curve are on the straight line between $\mathbb{E}_{P_{U,X}}[Y_0]$ and $\mathbb{E}_{P_{U,X}}[Y_1]$, this indicates that the benefit attributed to an increase in $\kappa$ arises solely from the incremental loosening of constraints, irrespective of who the constraints are placed upon, implying there is an absence of positive treatment effect heterogeneity. On the other hand, if some or all points (except $\Psi^F_{\kappa=0}(P_{U,X})$ and $\Psi^F_{\kappa=1}(P_{U,X})$) on the constraint-response curve fall above the line between $\mathbb{E}_{P_{U,X}}[Y_0]$ and $\mathbb{E}_{P_{U,X}}[Y_1]$, this means that the benefit attributed to an increase in $\kappa$ is due to a loosening of constraints \emph{and} there are subgroups of people who differentially benefit -- and the RC ODTR optimally allocates treatment based on this differential effect.  

A working model for the true, unknown curve can be represented by $\{m_\beta : \beta\}$; specifically, we opt to summarize the constraint-response curve using a linear working model: $m_{\beta}(\kappa) = \beta_0 + \beta_1\kappa$ \citep{robins2000marginal, petersen2014targeted, neugebauer2007nonparametric}, so that the causal quantity of interest is defined as
\begin{align}
&\argmin_{\beta}\mathbb{E}_{P_{U,X}}\left[\sum_{\kappa \in \bar{\kappa}}\left(\Psi^F_{\kappa}(P_{U,X}) - m_{\beta}(\kappa)\right)^2\right].
\label{MSM}
\end{align}
When treatment is likely to benefit all, we may be interested comparing this working MSM to the line connecting $\mathbb{E}_{P_{U,X}}[Y_0]$ and $\mathbb{E}_{P_{U,X}}[Y_1]$. If all points on the true constraint-response curve fall on the line between $\mathbb{E}_{P_{U,X}}[Y_0]$ and $\mathbb{E}_{P_{U,X}}[Y_1]$, then the working MSM will be equal to the true constraint-response curve, and thus $\beta_0$ on the working MSM will be equal to $\mathbb{E}_{P_{U,X}}[Y_0]$ and $\beta_1$ on the working MSM will be equal to $\mathbb{E}_{P_{U,X}}[Y_1] - \mathbb{E}_{P_{U,X}}[Y_0]$. Alternatively, if some or all points but $\Psi^F_{\kappa=0}(P_{U,X})$ and $\Psi^F_{\kappa=1}(P_{U,X})$ on the true constraint-response curve fall above the line between $\mathbb{E}_{P_{U,X}}[Y_0]$ and $\mathbb{E}_{P_{U,X}}[Y_1]$, then $\beta_0$ and/or $\beta_1$ on the working MSM will differ from $\mathbb{E}_{P_{U,X}}[Y_0]$ and/or $\mathbb{E}_{P_{U,X}}[Y_1] - \mathbb{E}_{P_{U,X}}[Y_0]$, respectively.  As explained in the previous paragraph, the latter scenario implies differential treatment benefit and thus utility of treatment allocation via the RC ODTR, while the former does not. Ultimately, we may be interested in testing the null hypothesis $H_0: \beta_0 = \mathbb{E}_{P_{U,X}}[Y_0]$ and $\beta_1 = \mathbb{E}_{P_{U,X}}[Y_1] - \mathbb{E}_{P_{U,X}}[Y_0]$ versus the alternative $H_A: \beta_0 \neq \mathbb{E}_{P_{U,X}}[Y_0]$ or $\beta_1 \neq \mathbb{E}_{P_{U,X}}[Y_1] - \mathbb{E}_{P_{U,X}}[Y_0]$ to test if such heterogeneity exists.


Finally, it could be of interest to understand the monetary savings or costs $C \in [0,\infty)$ incurred by administering the RC ODTR versus another rule $d$, such as the unconstrained optimal rule, relative to the clinical benefits or dropoff imposed by the RC ODTR. This translates to the following causal ICER \citep{o1994search, montoya2025cost}: $\frac{\mathbb{E}_{P_{U,X}}[C_{\tilde{g}^*_{\kappa}} - C_d]}{\mathbb{E}_{P_{U,X}}[Y_{\tilde{g}^*_{\kappa}} - Y_d]}$. In ADAPT-R, this would represent a cost-effectiveness measure of administering CCTs under the RC ODTR versus a comparator static rule, such as giving SOC to all.

\section{Identification, Estimation, and Inference}

\subsection{Identification}

We assume that our observed data were generated by sampling $n$ independent observations $O_i \equiv (W_i, A_i, Y_i)$, $i = 1, \ldots, n$, from a data generating system described by $\mathcal{M}^F$ above (e.g., the ADAPT-R study consists of 1,189 i.i.d. observations of $O$). 

As previously mentioned in \ref{odtr_noRC}, under the classic point-treatment randomization assumption ($Y_{a} \perp A|W$ for $a \in \{0,1\}$) and positivity assumption ($Pr\left(\min_{a \in \{0,1\}} g(a|W)>0\right) = 1$), which hold when treatment is randomized, as in ADAPT-R, the CATE can be identified as the blip $B_0(W)$. With these conditions, the following are identified: $S_0(\tau) = Pr(B_0(W) > \tau)$, $\eta_0 = \inf\{\tau : S_0(\tau) \leq \kappa\}$, and $\tau_0 = \max\{\eta_0,0\}$, so that the RC ODTR is identified as a function of the observed data distribution $P_0$ for each $\kappa \in \bar{\kappa}$; that is:
\begin{align*}
 \tilde{g}^*_{\kappa,0}(1|W) &=
  \begin{cases}
   \frac{\kappa - S_{0}(\tau_{0})}{Pr(B_0(W) = \tau_{0})},        & \text{if } B_0(W) = \tau_{0} \text{ and } \tau_{0} > 0 \\
   \mathbb{I}[B_0(W) > \tau_{0}],        & \text{otherwise.}
  \end{cases}
\end{align*}


Consider the mapping $\Psi_{\kappa}: \mathcal{M} \rightarrow \mathbb{R}$. For a given $\kappa$, $\Psi^F_{\kappa}(P_{U,X})$ is identified via the g-computation formula as follows \citep{robins1986new}:
\begin{align*}
\Psi_{\kappa}(P_0) &= \mathbb{E}_0\left[\sum_{a \in \{0,1\}}\mathbb{E}_0[Y|A=a,W]\tilde{g}^*_{\kappa,0}(a|W)\right].
\end{align*}
Equation (\ref{MSM}) is identified through the collection of g-computation formulas, $\left\{\Psi_{\kappa}(P_0):\kappa \in \bar{\kappa}\right\}$; that is:
\begin{align*}
    &\argmin_{\beta}\mathbb{E}_{0}\left[\sum_{\kappa \in \bar{\kappa}}\left(\Psi_{\kappa}(P_0) - m_{\beta}(\bar{\kappa})\right)^2\right].
\end{align*}
Under the above assumptions, the causal ICER is also identified; the following is its corresponding statistical parameter:
\begin{align*}
    &\frac{\mathbb{E}_0\left[\sum_{a \in \{0,1\}}\mathbb{E}_0\left[C|A=a,W\right]\tilde{g}^*_{\kappa,0}(a|W)\right] - \mathbb{E}_0\left[\mathbb{E}_0\left[C|A=d,W\right]\right]}{\Psi_{\kappa}(P_0) - \mathbb{E}_0\left[\mathbb{E}_0\left[Y|A=d,W\right]\right]}.
\end{align*}

\subsection{Estimation and Inference}

We denote estimators with a subscript $n$, so that, for example, an estimator of the true treatment mechanism $g_0$ is $g_n$. Estimates are functions of $P_n$, which is the empirical distribution that gives each observation weight $\frac{1}{n}$. Here, $\hat{\Psi}(P_n)$ is an estimate of the true parameter value $\Psi(P_0)$, by applying the estimator $\hat{\Psi}$ to an empirical distribution based on sampling from $P_0$.

\subsubsection{The RC ODTR and its Value}
The \texttt{SL.ODTR} package: https://github.com/lmmontoya/SL.ODTR can be used for estimating the RC ODTR and its value. To estimate the RC ODTR, one must first obtain an estimate of the blip function $B_n(W)$. This can be obtained via, for example, the optimal dynamic treatment rule SuperLearner \citep{montoya2022optimal, luedtkeSLODTR}, in which case the metalearner must be blip-based so that the output will contain an estimate of the blip function. Further, we recommend using the MSE as the risk function, since, here, our target is not directly predicting the binary treatment that maximizes the expected outcome, but instead estimating the most accurate blip values; thus, we choose the risk function that minimizes the squared error between the blip pseudo-outcome and predicted blip values. 

Then, for a range of $\tau$, one can estimate $S_0(\tau)$, $\eta_0$, and $\tau_0$ using their respective plug-in estimators: $S_n(\tau) = \frac{1}{n}\sum_{i=1}^n\mathbb{I}[B_n(W) > \tau]$, $\eta_n = \inf\{\tau : S_n(\tau) \leq \kappa\}$, and $\tau_n = \max\{\eta_n,0\}$. Finally, the RC ODTR can be estimated by:
\begin{align*}
 \tilde{g}^*_{\kappa,n}(1|W) &=
  \begin{cases}
   \frac{\kappa - S_{n}(\tau_{n})}{Pr(B_n(W) = \tau_{n})},        & \text{if } B_n(W) = \tau_{n} \text{ and } \tau_{n} > 0 \\
   \mathbb{I}[B_n(W) > \tau_{n}],        & \text{otherwise.}
  \end{cases}
\end{align*}

One targeted maximum likelihood estimator (TMLE) for estimating the value of the RC ODTR $\Psi_{\kappa}(P_0)$ can be implemented as follows. First, estimate nuisance parameters $\mathbb{E}_0[Y|A,W]$ and $g_0(A|W)$ using, for example, SuperLearner for prediction \citep{van2007super}, noting that if treatment is randomized (as in ADAPT-R), the true treatment mechanism could be used instead, but estimating it using a maximum likelihood estimate of the parameters of a correctly specified parametric model (such as a logistic regression regressing $A$ on $W$) can result in efficiency gains (see \cite{laan2003unified}, with an applied result in \cite{montoya2023efficient}). Call these estimates $\mathbb{E}_n[Y|A,W]$ and $g_n(A|W)$, respectively. Next, estimate the so-called ``clever" covariate $H_{n}(A,W) = \frac{\tilde{g}^*_{\kappa,n}(A|W)}{g_n(A|W)}$. Then, update the initial fit of $\mathbb{E}_n[Y|A,W]$  by running a logistic regression of $Y$ (which should be transformed between $0$ and $1$ if the outcome is continuous \citep{gruber2010targeted}) using the logit of $\mathbb{E}_n[Y|A,W]$ as offset and weights $H_{n}(A,W)$, with maximum likelihood estimation to estimate the intercept. Denote the predictions from this logistic regression as $\mathbb{E}^*_n[Y|A,W]$, from the updated fit. Finally, the TMLE of $\Psi_{\kappa}(P_0)$ is given by the plug-in estimator:
\begin{align*}
\hat{\Psi}_{\kappa}(P_n)&=\frac{1}{n} \sum_{i=1}^n  \sum_{a\in\{0,1\}} \mathbb{E}^*_n[Y|A_i=a,W_i]\tilde{g}^*_{\kappa,n}(A_i|W_i).
\end{align*}

The efficient influence function of $\Psi_{\kappa}(P_0)$ is:
\begin{align*}
    D_0(O) = & \frac{\tilde{g}^*_{\kappa,0}(A|W)}{g_0(A|W)}\left(Y - \mathbb{E}_0[Y|A,W]\right) + \left(\mathbb{E}_0[Y|A=1,W]\tilde{g}^*_{\kappa,0}(1|W) + \mathbb{E}_0[Y|A=0,W]\tilde{g}^*_{\kappa,0}(0|W)\right) \\
    & - \Psi_{\kappa}(P_0) - \tau_0(\tilde{g}^*_{\kappa,0}(1|W) - 
        \kappa).
\end{align*}
We refer the reader to \cite{luedtke2016statistical} for details on the consistency and regularity conditions needed for: 1) $\hat{\Psi}_{\kappa}(P_n)$ to be asymptotically linear, such that $\hat{\Psi}_{\kappa}(P_n) - \Psi_{\kappa}(P_0) = \frac{1}{n}\sum_{i=1}^nD_0(O_i) + o_{P_0}(n^{-1/2})$; 2) $\hat{\Psi}_{\kappa}(P_n)$ to be asymptotically efficient for $\Psi_{\kappa}(P_0)$, and 3) $\sigma_n^2 \equiv \frac{1}{n}\sum_{i=1}^nD(\tilde{g}^*_{\kappa,n}, \tau_n, P^*_n)(O_i)^2$ to be a consistent estimate for the variance of $D_0$, such that one can obtain 95\% confidence intervals:
\[\hat{\Psi}_{\kappa}(P_n) \pm \Phi^{-1}(0.975)\frac{\sigma_n}{\sqrt{n}}.\]

A layer of cross-fitting (i.e., cross-validated TMLE, or CV-TMLE) is recommended to estimate and obtain inference for estimates of $\Psi_{\kappa}(P_0)$ for two reasons: 1) to protect against the consequences of overfitting when estimating the nuisance parameters and the RC ODTR, and/or 2) if one is interested in estimating a so-called data adaptive parameter; we refer to \citep{montoya2022estimators, van2015targeted} for further details on the aforementioned points and implementation.

\subsubsection{Parameters of ``Constraint-Response" MSM Curve}
Next, we aim to estimate the parameters that summarize how the RC ODTR behaves as constraints are loosened on the proportion of people who can receive treatment in a population. Under the linear working marginal structural model specified in Equation \ref{MSM}, this can be done by regressing the collection of CV-TMLE estimates $\left\{\hat{\Psi}_{\kappa}(P_n):\kappa \in \bar{\kappa}\right\}$ on the vector of constraints of interest, $\bar{\kappa}$, using standard GLM software. The estimated intercept on this working MSM may be of particular interest, and inference on this parameter and its contrast with the average treatment effect can be obtained using the non-parametric bootstrap using 2.5\% and 97.5\% quantiles of the bootstrap distribution to construct confidence intervals.


\subsubsection{ICER}
Finally, to estimate the ICER, we refer the reader to \citep{montoya2025cost}. Briefly, one can use TMLE to estimate each of the components of the numerator and denominator. Then, using the functional delta method, one can derive the efficient influence function corresponding to the ICER parameter, to obtain influence-function based inference (such as confidence intervals).

\section{Application of the RC ODTR to the ADAPT-R Study}\label{adapt-res}

To recap, the overarching goal of the ADAPT-R study was to identify strategies for preventing lapses in HIV care. As part of a larger sequential multiple assignment randomized trial \citep{geng2023adaptive}, adults living with HIV were randomized to receive either CCTs or SOC, and a composite indicator of treatment success was considered as the outcome one year post-randomization. We estimated the RC ODTR on this sample to learn how to assign CCTs under constraints on the proportion of people who can receive CCTs in a population, prioritizing those who are predicted to \textit{most} likely benefit from CCTs, using the procedure outlined above. We note that, although the same SuperLearner configurations were used for the constrained rule as mentioned previously in the description of the unconstrained rule, the blip/CATE estimated for the constrained rule may have yielded a different blip/CATE estimate than the unconstrained rule due to random variations introduced by sample-splitting. We estimated the rule for the following range of constraints: $\bar{\kappa} = (0, 0.1, 0.2, ..., 0.9, 1.0)$. 


As expected by the design of the RC ODTR algorithm, the updated estimated blip/CATE cutoff for determining who should be treated (i.e., $\tau_n$, an estimate of $\tau_0$) increased progressively from 0 under the unconstrained rule as the constraints became tighter ($\tau_n$ = 0.06 at $\kappa = 0.9$ to $\tau_n$ = 0.27 at $\kappa = 0$). Accordingly, the percentage treated at each $\kappa$ was exactly $\kappa$ (so, for example, if the constraint was $\kappa = 0.1$, the RC ODTR recommended treatment to 10\% of the sample). This is unsurprising since the unconstrained optimal rule implied that all should receive CCTs (ie the estimated CATE was strictly positive); then, the RC ODTR algorithm should give the maximum amount of CCTs possible, which is what we see. Additionally, a probability of treatment (i.e., stochastic rule output), as opposed to a binary decision (i.e., deterministic rule output) was given to 4.79\%-36.59\% of participants, for $0.1 \leq \kappa \geq 0.9$ in $\bar{\kappa}$. This is also as expected, since the optimal rule was based on the indicator variables that did not uniformly split the CATE distribution into incremental buckets of 10\%, i.e., the pre-specified $\kappa$ values. 

Further, for each constraint, we estimated the expected counterfactual probability of treatment success under the RC ODTR using CV-TMLE. Results of this analysis are illustrated in Figure \ref{eyg_rc} and in Appendix B; these show that the probability of treatment success ranged from 66.55\% (95\% CI: 62.72\%, 70.37\%) at $\kappa = 0$ to 77.20\% (95\% CI: 73.91\%, 80.49\%) at $\kappa = 1$ for each $\kappa \in \bar{\kappa}$. Results additionally showed that any RC ODTR allocation rule with a constraint less than $\kappa = 1$ was statistically significantly less that giving CCTs to all (9.92\% difference for $\kappa = 0$ [95\% CI: 4.95\%, 14.90\%] and 1.63\% difference for $\kappa = .9$ [95\% CI: 0.47\%, 2.79\%]; noting we did not examine results for $0.9 < \kappa < 1$). Additionally, results showed that all RC ODTRs with a constraint greater than $\kappa = 0$ were statistically significantly better that giving SOC to all (1.63\% difference for $\kappa = .1$ [95\% CI: 0.28\%, 2.99\%] and 10.07\% difference for $\kappa = 1$ [95\% CI: 5.10\%, 15.05\%]; again, noting for this analysis we do not present results for $0 < \kappa < 0.1$). 

\begin{figure}[h]
    \centering
    \includegraphics[scale = .5]{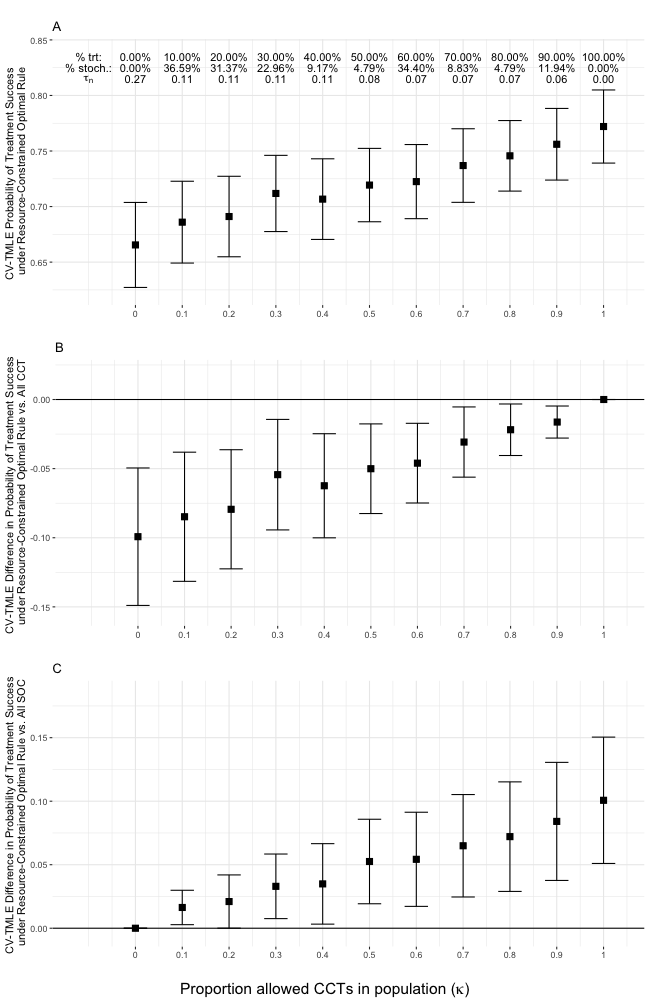}
    \caption{Cross-validated Targeted Maximum Likelihood Estimator (CV-TMLE) estimates of the value of the resource-constrained optimal dynamic treatment rule (RC ODTR; Panel A) compared to treating all with conditional cash transfers (CCTs; Panel B) and treating all with standard-of-care (SOC; Panel C) for different constraints on the proportion of people who can be treated in the population, $\kappa$. For Panel A, ``\% trt." and ``\% stoch." denote the percentage treated and the percentage given a stochastic treatment, respectively, with the RC ODTR; $\tau_n$ denotes the estimated updated (from 0) cutoff for deciding who to treat. For Panels B and C, black lines denote the null, i.e., no difference in contrasts.}
    \label{eyg_rc}
\end{figure}

Next, we examined if the probability of treatment success varies as a function of different resource constraints, $\kappa$ (results shown in Figure \ref{eyg_rc}). For a working MSM exploring the range  
\[\bar{\kappa}_{all} = (0, 0.1, 0.2, ..., 0.9, 1),\]
versus CV-TMLE estimates of the probability of treatment success, the estimated slope coefficient was $\hat{\beta_1}$ = 0.0948 ($2.5^{th}$ bootstrap quantile: 0.0419, $97.5^{th}$ bootstrap quantile: 0.1469). This means that for every for every 0.0948 increase in $\kappa$, the probability of treatment success increases by 0.95\% on average. The estimated intercept was $\hat{\beta_0}$ = 0.6720 ($2.5^{th}$ bootstrap quantile: 0.6492, $97.5^{th}$ bootstrap quantile: 0.7303). This line was not significantly different than the one that connects estimates of $\mathbb{E}_{P_{U,X}}[Y_0]$ and $\mathbb{E}_{P_{U,X}}[Y_1]$; the difference in slopes was -0.0117 ($2.5^{th}$ bootstrap quantile: -0.0246, $97.5^{th}$ bootstrap quantile: 0.0147) and the difference in intercepts was 0.0065 ($2.5^{th}$ bootstrap quantile: -0.0008, $97.5^{th}$ bootstrap quantile: 0.0387). From these results, there is insufficient evidence to conclude that the RC ODTR does better than the strategy that also allocates treatment based on constraints, but simply by chance (i.e., insufficient to reject the null hypothesis of no treatment effect heterogeneity).

\begin{figure}[h]
    \centering
    \includegraphics[scale = .65]{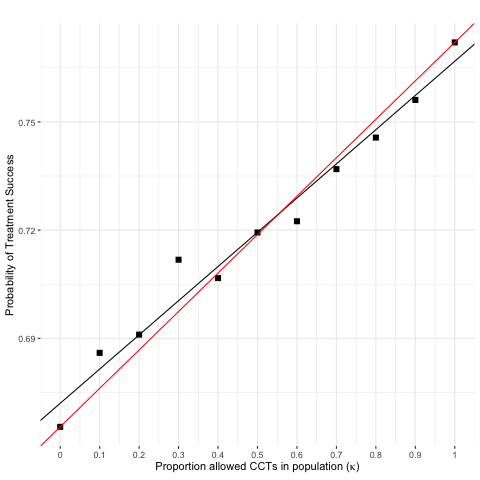}
    \caption{Working marginal structural model (MSM) of the probability of treatment success as a function of the proportion of people who can be treated in the population, $\kappa$. Squares are the cross-validated targeted maximum likelihood estimation (CV-TMLE) point estimates of the probability of treatment success for each $\kappa$. The black line represents the working model of the linear relationship between all $\kappa$ and their respective CV-TMLE estimates. The red line represents the line connecting estimates of $\mathbb{E}_{P_{U,X}}[Y_0]$ and $\mathbb{E}_{P_{U,X}}[Y_1]$. The extent to which the black line falls above the red represents the resource constrained optimal dynamic treatment rule's utility compared to allocating the allowed $\kappa$ of treatment at random.}
    \label{eyg_rc}
\end{figure}

\begin{figure}[h]
    \centering
    \includegraphics[scale = .55]{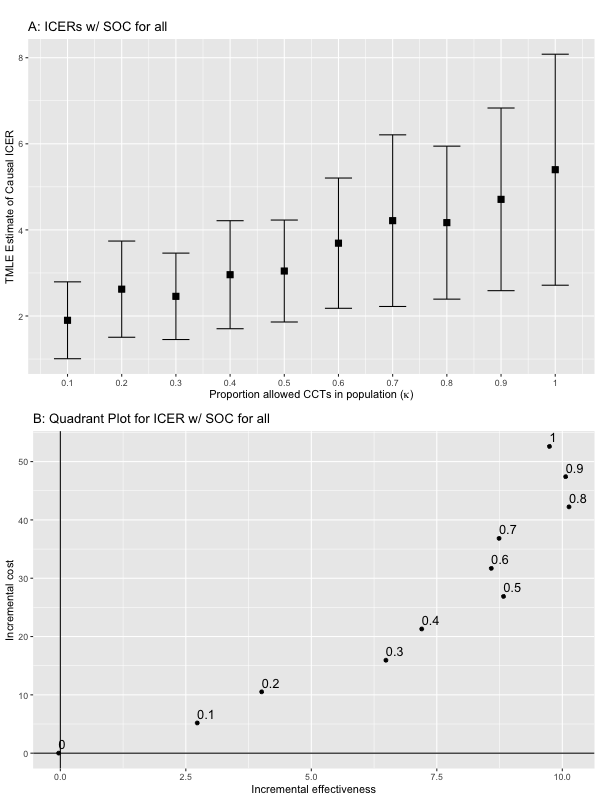}
    \caption{Incremental cost-effectiveness ratios (ICERs; Panel A) and quadrants (Panel B) comparing the cost effectiveness of the resource-constrained optimal dynamic treatment rule (RC ODTR) compared to the static rule of standard-of-care (SOC) for all. For Panel A, each square (with error bars denoting 95\% confidence intervals) depicts a point estimate of the causal ICER estimated with targeted maximum likelihood estimation (TMLE). For Panel B, each dot denotes a cost/effectiveness point for each constraint $\kappa$ (labeled above each dot). The $x$-axis displays the difference in probability of treatment success for the RC ODTR vs. SOC for all, for each constraint level $\kappa$ (i.e., effectiveness, the denominator for the ICER). The $y$-axis displays the difference in expected cost for the RC ODTR vs. SOC for all, for each constraint level $\kappa$ (i.e., cost, the numerator for the ICER).}
    \label{icer}
\end{figure}

Finally, we analyzed the ICERs of the RC ODTR at varying constraints compared to giving all persons SOC. For this, we estimated the causal ICER at each $\kappa$ value using TMLE; results for each of the ICERs and their components \citep{stinnett1998net} are shown in Figure \ref{icer} and Appendix B. Results show that the RC ODTR becomes progressively less cost effective as constraints are loosened, with the lowest ICER resulting from a resource constrained rule that treated the 10\% of the population most likely to benefit  (ICER for $\kappa$ = 0.1 is \$1.90 per additional person experiencing one-year treatment success [95\% CI; \$1.01, \$2.79]; ICER for $\kappa = 1$ is \$5.40 per additional person experiencing one-year treatment success [95\% CI; \$2.72, \$8.08]; difference = \$3.49 [95\% CI:  \$0.99, \$6.01]).

\section{Discussion}

In this paper, we described and presented an application of the RC ODTR, initially developed by \cite{luedtke2016statistical}, on trial data. In particular, we applied the RC ODTR and various methods of evaluating it to the ADAPT-R data. To the best of our knowledge, this is among the first illustrations of the RC ODTR approach. In addition, we described and presented several practical approaches for presenting the RC ODTR and its value (contrasts with other rules, an MSM, the causal ICER).  

We carried out these analyses to understand best ways of allocating CCTs, a behavioral intervention found to be effective (in expectation) among every subgroup defined by our covariates, among adults living with HIV in sub-Saharan Africa – given resource constraints in this setting. From our analyses, we learned that one strategy for preferentially allocating CCTs is a function of distance to clinic, working for wages, and self-employment – prioritizing those who walk $>$ 5 kilometers to clinic, are self-employed, and do not work for wages. We note that, while in this analysis the estimated unconstrained optimal allocation strategy yielded an interpretable rule, this may not necessarily be the case for a SuperLearner that includes uninterpretable (often more flexible) algorithms of the blip in its library.

Once constraints were placed on the proportion of patients who could receive CCTs in a population, there were statistically significant drop-offs in the probability of treatment success compared to the optimal treatment strategy of giving everyone CCTs at every constraint value. However, there were HIV care adherence benefits to administering CCTs to even a small percentage (e.g., 10\%) of the population compared to giving the care standard to all. MSM analyses showed that every additional loosening of constraints by 10\% translated to approximately a 1\% absolute increase in probability of treatment success. ICER analyses showed that the smallest ICER (most benefit for the least cost) occurred through offering the treatment to the 10\% of the population estimated to be the most likely to benefit, and supported the efficiency of this strategy rather than offering treatment to all. These results highlight the clinical importance of administering CCTs to a small pool of people. From this analysis, we learned there are both clinical and monetary benefits to doing so. 

Nevertheless, there was insufficient evidence of treatment effect heterogeneity significant enough to establish the RC ODTR as superior to a strategy that, like the RC ODTR, adheres to constraints on the proportion of people who can receive treatment in the population, but, unlike the RC ODTR, allocates treatment without regard to differential treatment effects. If there were stronger evidence of treatment effect heterogeneity, results from such an analysis could offer an alternative strategy for deploying CCTs that aims to preserve their effectiveness (because it allocates treatment to those who most benefit from it) while improving efficiency of their deployment (because it does not allocate treatment to those who are less likely to benefit from it) compared to strategies that are ``one-size-fits-all” --  a treatment-allocation feature that is crucial in resource-limited settings. In such cases, it may be useful to interrogate the true constraint response curve in a more flexible way using, for example, a non-linear data-adaptive MSM. That way, one can explore important features of the curve, such as where the curve is the steepest, to determine, for example, when loosening constraints is most worthwhile.

There are limitations of the presented RC ODTR and thus future directions that could be explored for the algorithm. For example, ADAPT-R is a sequential multiple assignment randomized trial \citep{kidwell2023sequential} with multi-level treatments randomized at different time-points based on patient responses. The results presented here were based on an RCT nested within ADAPT-R; results depended on a binary treatment given at a single time-point. While some researchers are exploring ways to address these limitations (see, e.g., \cite{caniglia2021estimating}), it would be of great interest in future work to consider SuperLearner-based estimation strategies for learning the resource-constrained optimal sequential strategy with multi-level treatments. Additionally, as \cite{luedtke2016statistical} also mention in their discussion, all results here are utilitarian-based -- they are purely focused on maximizing the expected outcome. Future modifications to this algorithm could maximize alternative objective functions or integrate additional constraints that move treatment allocation closer to distributive justice.

\section*{Acknowledgments}
Research reported in this publication was supported by NIH awards R01AI074345, K99/R00MH133985, and R01NR018801. The content is solely the responsibility of the authors and does not necessarily represent the official views of the NIH.

\bibliographystyle{plainnat}
\bibliography{mybib}

\appendix

\section{Appendix A}

Table \ref{app_table1} shows the estimated coefficients from the unconstrained optimal dynamic treatment rule SuperLearner. Any algorithms that begin with ``SL.blip.W\_" (i.e., algorithms 1-26) denote a main-terms, univariate generalized linear model (GLM) with only that covariate. For example, SL.blip.W\_age is a univariate GLM of a pseudo-outcome designed for blip estimation (as detailed in \cite{montoya2022optimal}) on age.

The unconstrained optimal dynamic treatment rule SuperLearner only placed weight on three such univariate GLMs with the following binary covariates: whether self-employed, whether work for a wage (casual work), and whether walk $>5$ kilometers to clinic.

\begin{table}[ht]
\centering
\begin{tabular}{rlr}
  \hline
 & Algorithm & Coefficient \\ 
  \hline
1 & SL.blip.W\_age & 0.00 \\ 
  2 & SL.blip.W\_audit\_cat & 0.00 \\ 
  3 & SL.blip.W\_CD4 & 0.00 \\ 
  4 & SL.blip.W\_distance\_clinic\_km & 0.00 \\ 
  5 & SL.blip.W\_educ & 0.00 \\ 
  6 & SL.blip.W\_fnua\_score & 0.00 \\ 
  7 & SL.blip.W\_housefarm & 0.00 \\ 
  8 & SL.blip.W\_marital & 0.00 \\ 
  9 & SL.blip.W\_mos\_score & 0.00 \\ 
  10 & SL.blip.W\_occup & 0.00 \\ 
  11 & SL.blip.W\_phq\_cat & 0.00 \\ 
  12 & SL.blip.W\_phq\_ind & 0.00 \\ 
  13 & SL.blip.W\_pregnant & 0.00 \\ 
  14 & SL.blip.W\_selfemploy & 0.37 \\ 
  15 & SL.blip.W\_sex & 0.00 \\ 
  16 & SL.blip.W\_time\_clinic\_min & 0.00 \\ 
  17 & SL.blip.W\_totaldiscspend & 0.00 \\ 
  18 & SL.blip.W\_totalfixed & 0.00 \\ 
  19 & SL.blip.W\_totalincome & 0.00 \\ 
  20 & SL.blip.W\_totalworkhrs & 0.00 \\ 
  21 & SL.blip.W\_wagework & 0.10 \\ 
  22 & SL.blip.W\_walk30 & 0.00 \\ 
  23 & SL.blip.W\_walk3k & 0.00 \\ 
  24 & SL.blip.W\_walk5k & 0.53 \\ 
  25 & SL.blip.W\_walk60 & 0.00 \\ 
  26 & SL.blip.W\_whostage & 0.00 \\ 
  27 & SL.mean & 0.00 \\ 
  28 & SL.glm & 0.00 \\ 
  29 & SL.bayesglm & 0.00 \\ 
  30 & SL.stepAIC & 0.00 \\ 
   \hline \\ \\
\end{tabular}
\caption{Estimated coefficients from the unconstrained optimal dynamic treatment rule SuperLearner.}  
\label{app_table1}
\end{table}

\section{Appendix B}

\begin{table}[ht]
\centering
\begin{tabular}{rlll}
  \hline
$\kappa$ & $\Psi_{\kappa}(P_{U,X})$ & $\Psi_{\kappa}(P_{U,X}) - \mathbb{E}_{P_{U,X}}[Y_1]$ & $\Psi_{\kappa}(P_{U,X}) - \mathbb{E}_{P_{U,X}}[Y_0]$ \\ 
  \hline
0.0 & 0.6655 [0.6272, 0.7037] & -0.0992 [-0.1490, -0.0495] & 0.0000 [0.0000, 0.0000] \\
0.1 & 0.6860 [0.6492, 0.7228] & -0.0848 [-0.1315, -0.0381] & 0.0163 [0.0028, 0.0299] \\
0.2 & 0.6910 [0.6548, 0.7273] & -0.0794 [-0.1225, -0.0363] & 0.0210 [0.0000, 0.0420] \\
0.3 & 0.7118 [0.6775, 0.7461] & -0.0544 [-0.0943, -0.0144] & 0.0330 [0.0076, 0.0584] \\
0.4 & 0.7067 [0.6704, 0.7430] & -0.0624 [-0.1001, -0.0247] & 0.0349 [0.0032, 0.0666] \\
0.5 & 0.7193 [0.6863, 0.7524] & -0.0501 [-0.0825, -0.0176] & 0.0525 [0.0193, 0.0858] \\
0.6 & 0.7225 [0.6891, 0.7558] & -0.0460 [-0.0749, -0.0172] & 0.0543 [0.0172, 0.0913] \\
0.7 & 0.7369 [0.7038, 0.7700] & -0.0308 [-0.0562, -0.0054] & 0.0649 [0.0246, 0.1052] \\
0.8 & 0.7457 [0.7139, 0.7774] & -0.0219 [-0.0405, -0.0032] & 0.0721 [0.0290, 0.1152] \\
0.9 & 0.7561 [0.7238, 0.7883] & -0.0163 [-0.0279, -0.0047] & 0.0841 [0.0376, 0.1306] \\
1.0 & 0.7720 [0.7391, 0.8049] & 0.0000 [0.0000, 0.0000] & 0.1007 [0.0510, 0.1505] \\
\hline \\ \\
\end{tabular}
\caption{Point estimates and estimated contrasts (with 95\% confidence intervals in brackets) of the counterfactual probability of treatment success under each resource constrained optimal rule.} 
\label{app_table2}
\end{table}

\begin{sidewaystable}[ht]
\centering
\begin{tabular}{rrrrlrrl}
  \hline
$\kappa$ & $\mathbb{E}_{P_{U,X}}[C_{\tilde{g}^*_{\kappa}} - C_1]$ & $\mathbb{E}_{P_{U,X}}[Y_{\tilde{g}^*_{\kappa}} - Y_1]$ & ICER: $\frac{\mathbb{E}_{P_{U,X}}[C_{\tilde{g}^*_{\kappa}} - C_1]}{\mathbb{E}_{P_{U,X}}[Y_{\tilde{g}^*_{\kappa}} - Y_1]}$ & $\mathbb{E}_{P_{U,X}}[C_{\tilde{g}^*_{\kappa}} - C_0]$ & $\mathbb{E}_{P_{U,X}}[Y_{\tilde{g}^*_{\kappa}} - Y_0]$ & ICER: $\frac{\mathbb{E}_{P_{U,X}}[C_{\tilde{g}^*_{\kappa}} - C_0]}{\mathbb{E}_{P_{U,X}}[Y_{\tilde{g}^*_{\kappa}} - Y_0]}$ \\ 
  \hline
0.00 & -52.60 & -9.78 & 5.38 [2.72, 8.04] & 0.01 & -0.03 & -0.16 [-0.58, 0.25] \\
0.10 & -47.42 & -7.02 & 6.76 [2.39, 11.12] & 5.18 & 2.73 & 1.90 [1.01, 2.79] \\
0.20 & -42.08 & -5.73 & 7.34 [2.08, 12.60] & 10.52 & 4.01 & 2.62 [1.51, 3.74] \\
0.30 & -36.67 & -3.26 & 11.25 [-2.82, 25.33] & 15.94 & 6.49 & 2.46 [1.45, 3.46] \\
0.40 & -31.30 & -2.54 & 12.30 [-5.49, 30.09] & 21.30 & 7.20 & 2.96 [1.70, 4.21] \\
0.50 & -25.72 & -0.92 & 28.11 [-76.99, 133.21] & 26.88 & 8.83 & 3.04 [1.86, 4.23] \\
0.60 & -20.91 & -1.16 & 18.04 [-24.20, 60.29] & 31.70 & 8.59 & 3.69 [2.18, 5.21] \\
0.70 & -15.76 & -1.00 & 15.71 [-24.05, 55.47] & 36.84 & 8.74 & 4.21 [2.22, 6.21] \\
0.80 & -10.36 & 0.39 & -26.61 [-177.55, 124.33] & 42.24 & 10.13 & 4.17 [2.39, 5.95] \\
0.90 & -5.19 & 0.32 & -16.07 [-82.63, 50.49] & 47.41 & 10.07 & 4.71 [2.59, 6.83] \\
1.00 & - & - & - & 52.60 & 9.74 & 5.40 [2.72, 8.08] \\
\hline \\ \\
\end{tabular}
\caption{Targeted maximum likelihood estimator estimates of the causal incremental cost effectiveness ratios (ICERs) and their components for each resource constraint $\kappa$ value (with 95\% confidence intervals in brackets). Units are dollars per additional person with one-year treatment success.} 
\end{sidewaystable}

\end{document}